\documentclass[twocolumn,showpacs,preprintnumbers,amsmath,amssymb]{revtex4}

\usepackage{graphicx}
\usepackage{dcolumn}
\usepackage{bm}
\usepackage{epsfig}

\begin{document}

\preprint{DESY~11--027\hspace{13.95cm} ISSN 0418--9833}
\preprint{February 2011\hspace{16.3cm}}

\title{\boldmath 
Ultrahigh-energy neutrino-nucleon deep-inelastic scattering and the
Froissart bound}

\author{Alexey~Yu.~Illarionov$^1$, Bernd A. Kniehl$^2$,
Anatoly~V.~Kotikov$^{2,3}$}
\affiliation{$^1$ Dipartimento di Fisica dell'Universit\'{a} di Trento,
via Sommarive 14, 38050 Povo, Trento, Italy\\
$^2$ {II.} Institut f\"ur Theoretische Physik, Universit\"at Hamburg,
Luruper Chaussee 149, 22761 Hamburg, Germany \\
$^3$ Bogoliubov Laboratory for
Theoretical Physics, JINR, 141980 Dubna (Moscow region), Russia}

\date{\today}

\begin{abstract}
We present a simple formula for the total cross section $\sigma^{\nu N}$ of
neutral- and charged-current deep-inelastic scattering of ultrahigh-energy
neutrinos on isoscalar nuclear targets, which is proportional to the structure
function $F_2^{\nu N}(M_V^2/s,M_V^2)$, where $M_V$ is the intermediate-boson
mass and $s$ is the square of the center-of-mass energy.
The coefficient in the front of $F_2^{\nu N}(x,Q^2)$ depends on the asymptotic
low-$x$ behavior of $F_2^{\nu N}$.
It contains an additional $\ln{s}$ term if $F_2^{\nu N}$ scales with a power of
$\ln(1/x)$.
Hence, an asymptotic low-$x$ behavior $F_2^{\nu N}\propto\ln^2(1/x)$, which is
frequently assumed in the literature, already leads to a violation of the
Froissart bound on $\sigma^{\nu N}$.
\end{abstract}


\maketitle

For more than a decade, large experiments have been searching for
ultrahigh-energy (UHE) cosmic neutrinos ($\nu$), with energies
$E_{\nu}>10^6$~GeV, using detectors scanning for events in large volumes of
water, ice, the Earth's atmosphere, and the lunar regolith
\cite{Becker:2007sv}.
While no clear indication of such an event has yet been reported, experimental
bounds on UHE-neutrino fluxes could be established, which, put together, now
cover energies way up to $10^{12}$~GeV
and start to constrain scenarios of
astrophysical interest.
Since these limits directly depend on the total cross section
$\sigma^{\nu N}(E_\nu)$ of UHE-neutrino deep-inelastic scattering (DIS) off
nucleons ($N$), it is an urgent task to provide reliable theoretical
predictions for the latter in the asymptotic high-$E_\nu$ regime, which lies
far beyond the one explored by laboratory-based $\nu N$ DIS experiments and
corresponds to asymptotically low values of Bjorken's scaling variable $x$.
This requires extrapolation over several orders of magnitude in $E_\nu$, for
which various approaches exist
\cite{Andreev:1979cp,McKay:1985nz,Gandhi:1998ri,Fiore:2003kc,%
Block:2006dz,Berger:2007ic,Block:2010ud}.
These are based on successful descriptions of the terrestrial data within the
framework of perturbative QCD and frequently impose the Froissart bound
\cite{Froissart:1961ux} on $\sigma^{\nu N}$. 
According to the latter, unitarity and analyticity limit the total cross
section of a scattering process not to grow faster with energy than
$\ln^2s$.

In this Letter, we derive a general formula for $\sigma^{\nu N}$ that is
remarkably concise and correctly accounts for the asymptotic high-energy
behavior making it perfectly suitable for UHE-neutrino phenomenology.
It is proportional to the DIS structure function $F_2^{\nu N}(x,Q^2)$, which
has a well-known representation in terms of parton distribution functions
(PDFs) within the parton model (PM) of QCD, with $x$ and the typical energy
scale $Q$ appropriately defined in terms of $E_\nu$ and $M_V$ ($V=W,Z$).
To be on the conservative side, we assume for the time being, as in
Refs.~\cite{Andreev:1979cp,McKay:1985nz,Gandhi:1998ri,Fiore:2003kc,%
Block:2006dz,Berger:2007ic,Block:2010ud}, that the available experimental data
on DIS allow for an extrapolation to very high (low) values of $E_\nu$ ($x$)
using an appropriate parameterization of $F_2^{\nu N}$.
If the latter rises too steeply as $x\to0$, then possible new QCD phenomena,
such as gluon saturation or recombination, color glass condensates, multiple
pomeron exchanges, etc., are expected to enter the stage as a cure at $x$
values below those currently probed by DIS experiments (for a review, see
Ref.~\cite{Reno:2004cx}).
We shall return to this issue below, considering two popular models of
screening \cite{GayDucati:2001hc,Bartels:2002cj}.
 
Specifically, we consider the charged-current (CC) and neutral-current (NC)
DIS processes,
\begin{eqnarray}
\nu(k)+N(P) & \to & \ell(k^\prime)+X,
\nonumber\\
\nu(k)+N(P) & \to & \nu(k^\prime)+X,
\label{eq:dis}
\end{eqnarray}
respectively, where $N=(p+n)/2$ denotes an isoscalar nucleon target of mass
$M$, $X$ collects the unobserved part of the final state, and the four-momentum
assignments are indicated in parentheses, and introduce the familiar kinematic
variables
\begin{equation}
s=(k+P)^2,\
Q^2=-q^2,\
x=\dfrac{Q^2}{2q\cdot P},\
y=\dfrac{q\cdot P}{k\cdot P},
\end{equation}
where $q=k-k^\prime$.
In the target rest frame, we have $s=M(2E_\nu+M)$ and $xy=Q^2/(2ME_\nu)$.
In the kinematic regime of interest here, the inclusive spin-averaged 
double-differential cross sections of processes~(\ref{eq:dis}) are, to very
good approximation, given by
\cite{McKay:1985nz}
\begin{equation}
\dfrac{d^2\sigma_i^{\nu N}}{dx\,dy} =
 \dfrac{G_F^2ME_{\nu}}{2\pi}K_i
 \left(\dfrac{M_{V}^2}{Q^2+M_V^2}\right)^2
K(y)F_2^{\nu N},
\label{eq:diff}
\end{equation}
where $i={\rm CC},{\rm NC}$, $G_F$ is Fermi's constant, and $K(y)=2-2y+y^2$.
In the so-called {\it wee parton} picture appropriate for the low-$x$ regime
\cite{Berger:2007ic}, we have
$K_{\rm CC}=1$ and $K_{\rm NC}=1/2-x_w+(10/9)x_w^2$, where
$x_w=\sin^2\theta_w$, with $\theta_w$ being the weak mixing angle.
Using $x_w=0.231$ \cite{Nakamura:2010zzi}, we have $K_{\rm NC}=0.328$.
The contributions due to the structure functions $F_L^{\nu N}$ and
$F_3^{\nu N}$ to the r.h.s.\ of Eq.~(\ref{eq:diff}) are negligibly small in our
applications;
$F_L^{\nu N}$ tends to zero as $Q^2$ rises \cite{Kotikov:1993yw}, and
$F_3^{\nu N}$ essentially refers to valence partons, which hardly contribute in
the low-$x$ regime.

\begin{figure}
\begin{center}
\epsfig{figure=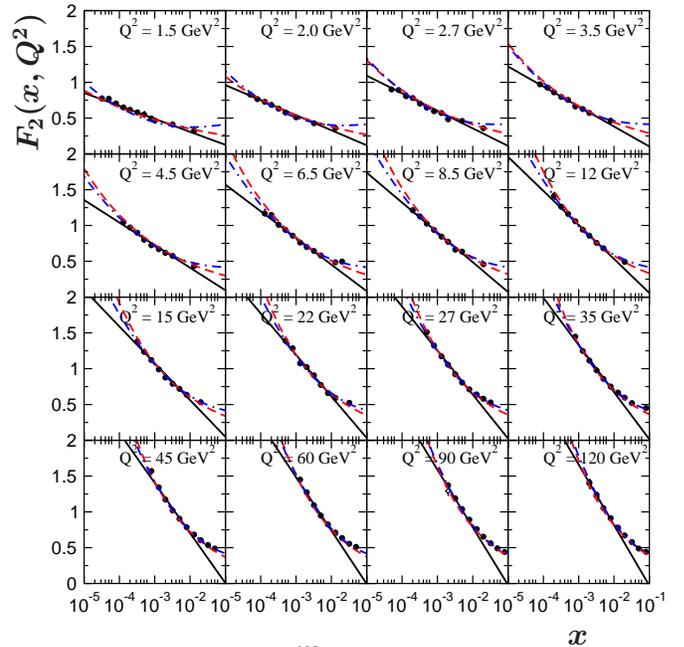,width=1.0\linewidth}
\end{center}
\vspace{-0.75cm}
\caption{Measurement of $F_2^{\ell N}$ \cite{:2009wt} compared with our fit using
the improved H ansatz as well as the PM and BBT results.}
\label{Fig1}
\end{figure}

Detailed inspection of the available $\ell N$ DIS data (see, e.g.,
Fig.~\ref{Fig1} for $ep$ data from HERA~I \cite{:2009wt}) suggests that, in
the limit $x \to 0$, $F^{\ell N}_2$ exhibits a singular behavior of the form
$F^{\ell N}_2(x,Q^2)\simeq x^{-\delta}\tilde{F}^{\ell N}_2(x,Q^2)$, where 
$\delta$ is a small positive number and $\tilde{F}^{\ell N}_2$
diverges less strongly than any power of $x$, {\it i.e.},
$\tilde{F}^{\ell N}_2(x,Q^2)/x^{-\lambda}\to0$ as $x\to0$ for any positive
number $\lambda$.
Assuming a symmetric quark sea, as is appropriate for the low-$x$ regime, we
have $F_2^{\nu N}(x,Q^2)=(18/5)F_2^{\ell N}(x,Q^2)$, so that the low-$x$
behavior of $\tilde F^{\ell N}_2$ carries over to $\tilde F^{\nu N}_2$.

Imposing the lower cut-off $Q_0^2$ on $Q^2$, the total cross sections of
processes~(\ref{eq:dis}) are evaluated as
\begin{equation}
\sigma^{\nu N}_i(E_\nu)=
	\dfrac{1}{2ME_{\nu}}\int_{Q_0^2}^{2ME_\nu}dQ^2
	\int_{\hat{x}}^{1}\dfrac{dx}{x}\,\dfrac{d^2\sigma^{\nu N}_i}{dx\,dy},
\label{eq:tot}
\end{equation}
where $\hat x=Q^2/(2ME_\nu)$.
Inserting Eq.~(\ref{eq:diff}), Eq.~(\ref{eq:tot}) becomes
\begin{eqnarray}
\sigma^{\nu N}_i(E_\nu)&=&
	\dfrac{G_F^2}{4\pi} K_i 
	\int_{Q_0^2}^{2ME_\nu}dQ^2
	\left(\dfrac{M_V^2}{Q^2+M_V^2}\right)^2
\nonumber\\
	&&\times
	\int_{\hat{x}}^{1}\dfrac{dx}{x}
        K\left(\dfrac{\hat{x}}{x}\right)
	F_2^i(x,Q^2).
\label{eq:tot1}
\end{eqnarray}
The inner integral on the r.h.s.\ of Eq.~(\ref{eq:tot1}) can be rewritten as
the Mellin convolution $K(\hat{x})\otimes F^{\nu N}_2(\hat{x},Q^2)$.
Exploiting the low-$x$ asymptotic form
$F^{\nu N}_2(x,Q^2)\simeq x^{-\delta}\tilde F^{\nu N}_2(x,Q^2)$
explained above, this Mellin transform may be represented, at small values of
$\hat{x}$, in the factorized form
$\tilde M(\hat{x},Q^2,1+\delta)F_2^{\nu N}(\hat{x},Q^2)$ up to terms of
${\cal O}(\hat{x})$ \cite{Lopez:1979bb}.
Here,
\begin{eqnarray}
&&\tilde M(\hat{x},Q^2,1+\delta)=
2\left(\dfrac{1}{\tilde{\delta}(\hat{x},Q^2)}-\dfrac{1}{\delta}\right)
+M(1+\delta),\quad
\\
&&\dfrac{1}{\tilde{\delta}(x,Q^2)}=
	\dfrac{1}{\tilde F^{\nu N}_2(x,Q^2)}
	\int_x^1 \dfrac{dy}{y}\tilde{F}^{\nu N}_2(y,Q^2),
\label{eq:delta}
\\
&&M(n) = \int_0^1 dx \, x^{n-2} K(x) =
	\dfrac{2}{n-1}-\dfrac{2}{n}+\dfrac{1}{n+1}.
\end{eqnarray}
Hence, Eq.~(\ref{eq:tot1}) becomes
\begin{eqnarray}
\sigma^{\nu N}_i(E_\nu)&\simeq&
	\dfrac{G_F^2}{4\pi} K_i 
	\int_{Q_0^2}^{2ME_\nu}dQ^2
	\left(\dfrac{M_V^2}{Q^2+M_V^2}\right)^2
\nonumber\\
	&&\times
\tilde M(\hat{x},Q^2,1+\delta)F_2^{\nu N}(\hat{x},Q^2).
\label{eq:tot2}
\end{eqnarray}
Because the $Q^2$ dependence of $F_2^{\nu N}(\hat{x},Q^2)$ and hence
$\tilde M(\hat{x},Q^2,1+\delta)$ is only logarithmic, the factor
$[M_V^2/(Q^2+M_V^2)]^2$ essentially fixes the scale $Q^2=M_V^2$
\cite{Fiore:2003kc}, so that Eq.~(\ref{eq:tot2}) simplifies to
\begin{equation}
\sigma^{\nu N}_i(E_\nu)\simeq
	\dfrac{G_F^2}{4\pi} K_i M_V^2
\tilde M(\tilde{x},M_V^2,1+\delta)F_2^{\nu N}(\tilde{x},M_V^2),
\label{eq:tot3}
\end{equation}
where $\tilde x=M_V^2/(2ME_\nu)$.
This is our master formula.
Further simplification depends on the actual size of $\delta$, and we
distinguish two cases.
(1) If $\delta$ is not too small, so that $\hat{x}^{\delta}\ll\text{const}$,
then the lower limit $\hat{x}$ of the inner integral on the r.h.s.\ of
Eq.~(\ref{eq:tot1}) may be put to zero, so that
\begin{equation}
\tilde M(\hat{x},Q^2,1+\delta)=M(1+\delta)=
\dfrac{4+3\delta+\delta^2}{\delta(\delta+1)(\delta+2)}
\label{eq:largedelta}
\end{equation}
becomes independent of $\hat{x}$ and $Q^2$.
(2) On the other hand, if $\delta\ll1$, then we have
\begin{equation}
\tilde M(\hat{x},Q^2,1+\delta)=\tilde M(\hat{x},Q^2,1)
=\dfrac{2}{\tilde{\delta}(\hat{x},Q^2)}-\dfrac{3}{2}.
\label{eq:smalldelta}
\end{equation}
We note that $\tilde{\delta}$ is determined by the asymptotic low-$x$ behavior
of $\tilde F^{\nu N}_2$.
For instance, if $\tilde F^{\nu N}_2(x,Q^2)\propto\ln^p(1/x)$ for $x\to0$, then
$1/\tilde{\delta}(x,Q^2)=\ln(1/x)/(p+1)$ \cite{Kotikov:1998qt}.

\begin{table}[t]
\begin{center}
\begin{tabular}{|cccccc|}
\hline
$i$ & $c_i\times10^{3}$ & $\delta_i\times10^{2}$ & $a_{1i}\times10^{2}$
& $a_{2i}\times10^{3}$ & $b_{i}\times10^{2}$ \\
\hline
 0 & $189.4$ & $10.90$ & $-8.471$ & $12.92$ & $2.689$ \\
 1 & $1.811$ & $6.249$ & $4.190$ & $0.2473$ & $11.63$ \\
 2 & $-0.6054$ & $-0.3722$ & $-0.3976$ & $1.642$ & $-0.7307$ \\
\hline
\end{tabular}
\end{center}
\caption{%
The values of the fit parameters appearing in Eqs.~(\ref{CTEQ6:C}),
(\ref{CTEQ6:d}), (\ref{BBT:F2.1}), and (\ref{Haidt:F2a}).}
\label{Tab:CTEQ6}
\end{table}

Now we apply Eq.~(\ref{eq:tot3}) to the three most popular types of
$F_2^{\ell N}$ parameterization, namely
the standard PM representation implemented with up-to-date proton PDFs
\cite{:2009wt,Martin:2009bu,Lai:2010vv}, a modification of the impressively
simplistic log-log form proposed by Haidt (H) \cite{Haidt:1999ps}, and the more
sophisticated form recently introduced by Berger, Block, and Tan (BBT)
\cite{Block:2006dz}.
While the $Q^2$ dependence of the PM representation of $F_2^{\nu N}$ is
governed by the DGLAP evolution, those of the heuristic H and BBT forms are
directly determined by global fits to experimental data covering a wide $Q^2$
range.
In the low-$x$ regime, the PM parameterization of $F^{\ell N}_2$ may be
well approximated by the following ansatz: 
\begin{equation}
F^{\ell N}_{2,\text{PM}}(x,Q^2)=C_\text{PM}(Q^2)x^{-\delta_\text{PM}(Q^2)},
\label{n8.1}
\end{equation}
with
\begin{eqnarray}
C_\text{PM}(Q^2) &=& c_0 + c_1 \ln{Q}^2 + c_2 \ln^2{Q}^2,
\label{CTEQ6:C}\\
\delta_\text{PM}(Q^2) &=& \delta_0 + \delta_1 \ln{Q}^2 + \delta_2 \ln^2{Q}^2, 
\label{CTEQ6:d}
\end{eqnarray}
where it is understood that $Q^2$ is taken in units of GeV${}^2$.
To suppress higher-twist effects, impose the cut $Q^2>3.5$~GeV$^2$.
Fitting Eqs.~(\ref{n8.1})--(\ref{CTEQ6:d}) to the result for $F_2^{\ell N}$
evaluated at next-to-leading order (NLO) with the HERAPDF1.0 \cite{:2009wt}
proton PDFs, we obtain the values of $c_i$ and $\delta_i$ collected in
Table~\ref{Tab:CTEQ6}. 
From Eq.~(\ref{CTEQ6:d}) and Table~\ref{Tab:CTEQ6}, we glean that
\begin{equation} 
\delta_\text{PM}(M_Z^2)\approx\delta_\text{PM}(M_W^2)\approx0.37,
\label{eq:cteqdelta}
\end{equation}
so that Eq.~(\ref{eq:tot3}) is to be used with Eq.~(\ref{eq:largedelta}).
Using the MSTW \cite{Martin:2009bu} and CT10 \cite{Lai:2010vv} PDFs, we obtain
$\delta_\text{PM}(M_V^2)\approx0.35$ and 0.38, respectively.
The resulting high-$E_\nu$ behavior
$\sigma^{\nu N}_i(E_\nu)\propto \tilde{x}^{-\delta_\text{PM}(M_V^2)}$
is in good agreement with other studies \cite{Gandhi:1998ri}. 

For the reader's convenience, we recollect here the BBT parameterization of
$F_2^{\ell N}$ appropriate for the range $x<x_P=0.11$ \cite{Block:2010ud}
relevant for our applications \footnote{%
We verified that the contribution to Eq.~(\ref{eq:tot1}) from
$F^{\ell N}_{2,\text{BBT}}$ in the range $x_P<x<1$ is numerically
insignificant, in agreement with
Refs.~\cite{Block:2006dz,Berger:2007ic,Block:2010ud}.}. 
It reads \cite{Block:2006dz,Berger:2007ic,Block:2010ud}:
\begin{eqnarray}
F^{\ell N}_{2,\text{BBT}}(x,Q^2) &=& (1-x)
	\left[A_0 + A_1(Q^2) \ln\dfrac{x_P(1-x)}{x(1-x_P)}
\right.
\nonumber\\
&&{}+\left.A_2(Q^2) \ln^2\dfrac{x_P(1-x)}{x(1-x_P)}\right],
\label{BBT:F2}
\end{eqnarray}
where $A_0=F_P/(1-x_P)$, with $F_P=0.413$ \cite{Block:2010ud}, and  
\begin{equation}
A_i(Q^2)=a_{i0} + a_{i1} \ln{Q^2} + a_{i2} \ln^2{Q^2}\quad(i=1,2),
\label{BBT:F2.1}
\end{equation}
with the values of $a_{ij}$ listed in Table~\ref{Tab:CTEQ6}
\cite{Block:2010ud}.
Here, Eq.~(\ref{eq:tot3}) is to be used with Eq.~(\ref{eq:smalldelta}), and
we find
\begin{eqnarray}
\dfrac{1}{\tilde{\delta}_\text{BBT}(x,Q^2)}\!&\!\simeq\!&\!
	\dfrac{\sum_{i=0}^2A_i\ln^{i+1}(x_P/x)/(i+1)}
	      {\sum_{i=0}^2A_i\ln^i(x_P/x)}
\simeq \dfrac{1}{3} \ln\dfrac{x_P}{x}.
\nonumber\\
\label{BBT:d}
\end{eqnarray}
From Eqs.~(\ref{BBT:F2}) and (\ref{BBT:d}), we glean that, in the high-energy
limit $s\to\infty$, 
$F^{\ell N}_{2,\text{BBT}}(\tilde{x},M_V^2)\propto\ln^2s$ and
$1/\tilde{\delta}_\text{BBT}(\tilde{x},M_V^2)\propto\ln s$.
This leads us to the important observation that
$\sigma_{\text{BBT}}^{\nu N}\propto\ln^3s$, which manifestly violates the
Froissart bound \cite{Froissart:1961ux} in contrast to what is stated in
Refs.~\cite{Block:2006dz,Berger:2007ic,Block:2010ud}.
This violation of the Froissart bound is attributed to the presence of the
$\ln^2x$ term in Eq.~(\ref{BBT:F2}).
On the other hand, if $F_2^{\ell N}$ just rises linearly in $\ln x$ as
$x\to0$, then $\sigma_i^{\nu N}\propto\ln^2s$ in accordance with the Froissart
bound.

In fact, this is the case for the original H ansatz \cite{Haidt:1999ps},
$B\ln(x_0/x)\ln(1+Q^2/Q_0^2)$, which contains just three fit parameters.
To enable the fit quality to be improved, we introduce three more parameters by
writing
\begin{eqnarray}
F^{\ell N}_{2,\text{H}}(x,Q^2) &=& B_0 + B_1(Q^2) \ln \dfrac{x_0}{x},
\nonumber\\
B_1(Q^2)&=&\sum_{i=0}^2 b_i \ln^i \left(1+\dfrac{Q^2}{Q^2_0}\right).
\label{Haidt:F2a}
\end{eqnarray}
Eq.~(\ref{eq:tot3}) is again to be used with Eq.~(\ref{eq:smalldelta}), and
we obtain
\begin{equation}
\dfrac{1}{\tilde{\delta}_\text{H}(x,Q^2)} \simeq
\dfrac{\sum_{i=0}^1B_i\ln^{i+1}(x_0/x)/(i+1)}
	      {\sum_{i=0}^1B_i\ln^i(x_0/x)}
\simeq \dfrac{1}{2} \ln\dfrac{x_0}{x},
\label{Haidt:d}
\end{equation}
so that $\sigma_\text{H}^{\nu N}\propto\ln^2s$ as it should.
Fitting Eq.~(\ref{Haidt:F2a}) to the recent combination of the complete H1 and
ZEUS data sets on $F_2^{\ell N}$ from HERA~I \cite{:2009wt} with the cuts
$x<0.01$ and 
$Q^2>1.5$~GeV$^2$,
we obtain
$x_0=0.05791$, $Q^2_0=2.578$~GeV${}^2$, $B_0=0.1697$, and the values of
$b_i$ listed in Table~\ref{Tab:CTEQ6}, with
$\chi^2/\text{d.o.f.}=422/175\approx2.41$.

Looking at Fig.~\ref{Fig1}, we observe that our fit (solid lines) indeed
yields a surprisingly good description of the experimental data over the full
$x$ and $Q^2$ ranges considered.
The approximation works particularly well for low $x$ and large $Q^2$ values
and is thus likely to allow for a reliable extrapolation to the $x$ and $Q^2$
ranges relevant for UHE-neutrino physics.
In fact, switching to the cut $x<10^{-3}$ reduces $\chi^2/\text{d.o.f.}$ by
roughly a factor of three, to $\chi^2/\text{d.o.f.}=58/69 \approx 0.84$.
For comparison, also the PM results evaluated from
Eqs.~(\ref{n8.1})--(\ref{CTEQ6:d}) (dashed lines) and the BBT results
evaluated from Eqs.~(\ref{BBT:F2}) and (\ref{BBT:F2.1}) (dash-dotted line),
which are hardly distinguishable from one another, are shown in
Fig.~\ref{Fig1}.

\begin{figure}
\begin{center}
\epsfig{figure=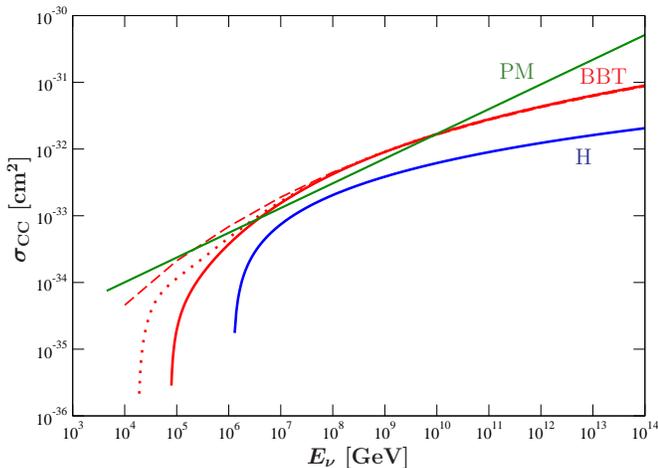,width=1.0\linewidth}
\end{center}
\vspace{-0.75cm}
\caption{Predictions for $\sigma_\text{CC}^{\nu N}(E_\nu)$ evaluated from the PM,
BBT, and H parameterizations of $F_2^{\ell N}(x,Q^2)$.
In the BBT case, also the improved high-$E_\nu$ approximation and the exact
evaluation are shown.}
\label{Fig2}
\end{figure}

We now consider $\nu N$ DIS with UHE neutrinos.
For the sake of brevity, we focus our attention on CC DIS.
The corresponding NC results may be obtained by substituting
$K_\text{CC}\to K_\text{NC}$ and $M_W\to M_Z$ in our formulas.
In order to determine the range of validity of our master formula
(\ref{eq:tot3}) for $\sigma_\text{CC}^{\nu N}$, we compare it with the exact
formula (\ref{eq:tot1}), which requires two-dimensional numerical integration,
for the PM, BBT, and H cases considered above.
In each case, we find excellent agreement for $E_\nu$ values of order
$10^7$~GeV and above, which corresponds to $x$ values of order $10^{-3}$ and
below in $F_2^{\ell N}$.
This is illustrated for the BBT case in Fig.~\ref{Fig2}, where
the approximate evaluation of Eq.~(\ref{eq:tot3}) with
Eqs.~(\ref{eq:smalldelta}) and (\ref{BBT:d}) is compared with the exact one of
Eq.~(\ref{eq:tot1}) with Eqs.~(\ref{BBT:F2}) and (\ref{BBT:F2.1}) (dashed
line).
The large-$E_\nu$ approximation may be somewhat improved by evaluating
$\tilde{\delta}(x,Q^2)$ by one-dimensional integration via Eq.~(\ref{eq:delta})
instead of using Eq.~(\ref{BBT:d}) (dotted line). 

The PM and H results for $\sigma_\text{CC}^{\nu N}$ evaluated from our
master formula (\ref{eq:tot3}), with Eqs.~(\ref{eq:largedelta}) and
(\ref{eq:cteqdelta}) in the PM case and with Eqs.~(\ref{eq:smalldelta}) and
(\ref{Haidt:d}) in the H case, are also displayed in Fig.~\ref{Fig2}.
Comparing them with the corresponding BBT result, we observe that all three
predictions agree relatively well in the range
$10^7$~GeV${}\alt E_\nu\alt10^9$~GeV, where the high-$E_\nu$ approximation is
already working and the respective $F_2^{\ell N}$ parameterizations are still
constrained by the HERA data.
However, these three predictions steadily diverge as $E_\nu$ further increases
until they differ by 1--2 orders of magnitude at typical UHE values of $E_\nu$,
reflecting the different low-$x$ behaviors of the respective parameterizations
of $F_2^{\ell N}$.

In summary, we derived a novel concise relationship, given by
Eqs.~(\ref{eq:tot3})--(\ref{eq:smalldelta}), between the total cross section
$\sigma_i^{\nu N}(E_\nu)$ of CC and NC $\nu N$ DIS in the high-$E_\nu$ limit
and the structure function $F_2^{\ell N}(x,Q^2)$ in the low-$x$ limit.
It is particularly useful for applications to UHE-neutrino physics providing
reliable predictions in a very quick and convenient way as it does.
Being given in terms of a closed analytic formula, it also allows one to
unambiguously determine if $\sigma_i^{\nu N}$ resulting from a given
functional form of $F_2^{\ell N}$ satisfies the Froissart bound
\cite{Froissart:1961ux} or not, while this is hardly possible using the
numerical solution of the two-dimensional integral in Eq.~(\ref{eq:tot1}).
Specifically, if $F_2^{\ell N}$ exhibits a low-$x$ behavior
${}\propto\ln^p(1/x)$, which corresponds to a high-$s$ behavior
${}\propto\ln^ps$ in Eq.~(\ref{eq:tot3}), then the coefficient $\tilde{M}$ in
that equation produces an additional factor ${}\propto\ln s$, so that the
Froissart bound is violated for $p>1$.
In fact, this is the case for the BBT
\cite{Block:2006dz,Berger:2007ic,Block:2010ud} parameterization of
$F_2^{\ell N}$, for which $p=2$.
On the other hand, the H \cite{Haidt:1999ps} one is characterized by $p=1$, so
that the Froissart bound is satisfied.
This motivated us to update the analysis of Ref.~\cite{Haidt:1999ps} by fitting
our improved ansatz~(\ref{Haidt:F2a}) to the recent combination of the complete
H1 and ZEUS data on $F_2^{\ell N}$ from HERA~I \cite{:2009wt}.

For completeness, we also performed a fit to the PM result for $F_2^{\ell N}$
evaluated at NLO with an up-to-date set of proton PDFs, namely the
HERAPDF1.0 one obtained by H1 and ZEUS by fitting their own data
\cite{:2009wt}, and presented the resulting prediction for $\sigma_i^{\nu N}$.
As expected, the low-$x$ behavior of the PM result for $F_2^{\ell N}$ is
too singular for $\sigma_i^{\nu N}$ to satisfy the Froissart bound.
It is likely that the inclusion of nonlinear terms, such as screening
corrections generated by gluon saturation or recombination, in the evolution
equations will cure this problem \cite{Reno:2004cx}.
In fact, considering the Ayala--Gay-Ducati--Levin \cite{GayDucati:2001hc} and
the generalized Golec-Biernat--W\"usthoff (GBW) \cite{Bartels:2002cj} models of
saturation, where, due to their specific gluon densities,
$F_2^{\ell N}\propto Q^2\ln^p(1/x)$ with $p=1$ and $p=0$, respectively, we
obtain $\sigma_i^{\nu N}\propto\ln^{p+2} s$, where the second additional
logarithm arises from the $Q^2$ integration in Eq.~(\ref{eq:tot2}).
Thus, saturation strongly modifies the power-like perturbative asymptotics of
total cross sections and has the potential to restore the Froissart bound, as
in the case of the GWB model.

Future measurements of $\nu N$ DIS with UHE neutrinos will eventually provide
direct access to the low-$x$ asymptotic behavior of $F_2^{\ell N}$, far beyond
the reach of accelerator experiments, and our new relationship will provide a
convenient tool to expose it.
On the theoretical side, one important lesson to be learned from our specific
example, where total cross sections could be simply related to structure
functions in the framework of perturbation theory, is that the direct
application of the Froissart bound to structure functions represents a
potential pitfall, of which we wish to caution the reader.

We thank Jochen Bartels, Lev Lipatov, and G\"unter Sigl for useful discussions.
This work was supported in part by BMBF Grant No.\ 05H09GUE and HGF Grant
No.\ HA~101.
The work of A.V.K. was supported in part by DFG Grant No.\ INST 152/465--1,
Heisenberg-Landau Grant No.~5, and RFBR Grant No.\ 10--02--01259--a.

\end{document}